\documentclass[apj]{emulateapj}
\usepackage{apjfonts}
\usepackage{epstopdf}
\usepackage{amsmath}

\def\wig#1{\mathrel{\hbox{\hbox to 0pt{%
          \lower.5ex\hbox{$\sim$}\hss}\raise.4ex\hbox{$#1$}}}}

\shorttitle{Cool Brown Dwarf Variability}

\newcommand{\teff}{$T_{\rm eff}$}

\newcommand{\cp}{\citep}
\newcommand{\ct}{\citet}

\newcommand{\fsed}{$f_{\rm sed}$} 
\newcommand{\nas}{Na$_2$S}

\slugcomment{Accepted for publication in ApJ Letters}

\begin{document}

\title{Spectral Variability from the Patchy Atmospheres of T and Y Dwarfs }

\author{Caroline V. Morley\altaffilmark{1}, Mark S. Marley\altaffilmark{2},  Jonathan J. Fortney\altaffilmark{1}, Roxana Lupu\altaffilmark{2}}

\altaffiltext{1}{Department of Astronomy and Astrophysics, University of California, Santa Cruz, CA 95064; cmorley@ucolick.org}
\altaffiltext{2}{NASA Ames Research Center} 

\begin{abstract}
Brown dwarfs of a variety of spectral types have been observed to be photometrically variable. Previous studies have focused on objects at the L/T transition, where the iron and silicate clouds in L dwarfs break up or dissipate. However, objects outside of this transitional effective temperature regime also exhibit variability. Here, we present models for mid-late T dwarfs and Y dwarfs. We present models that include patchy salt and sulfide clouds as well as water clouds for the Y dwarfs. We find that for objects over 375 K, patchy cloud opacity would generate the largest amplitude variability within near-infrared spectral windows. For objects under 375 K, water clouds also become important and generate larger amplitude variability in the mid-infrared. We also present models in which we perturb the temperature structure at different pressure levels of the atmosphere to simulate hot spots. These models show the most variability in the absorption features between spectral windows. The variability is strongest at wavelengths that probe pressure levels at which the heating is the strongest. The most illustrative types of observations for understanding the physical processes underlying brown dwarf variability are simultaneous, multi-wavelength observations that probe both inside and outside of molecular absorption features.

\end{abstract}

 
\section{Introduction}

	Brown dwarfs, the lowest-mass product of star formation, lack sustained hydrogen fusion and cool continuously, passing through the same temperature ranges as planets. Easier to observe than exoplanets, they are the first extrasolar substellar objects on which we have observed weather on other worlds, creating time-varying spectral features. 

Clouds form in brown dwarfs of most spectral types; if regionally heterogeneous, they cause photometric variability as cloudier hemispheres rotate in and out of view. L dwarf clouds are dusty layers of iron and silicates \cp{Tsuji96, Allard01, Marley02, Burrows06, Cushing08}. At the L/T transition, these clouds form holes or dissipate, leaving the early T dwarfs relatively cloud-free \cp{AM01, Burgasser02, Kirkpatrick05}. In the mid-late T dwarfs, alkali salts and sulfides solidify, reddening late T dwarfs which are otherwise quite blue in the near-infrared \cp{Lodders99, Visscher06, Morley12}. 

In the coolest brown dwarfs, the Y dwarfs, volatile species condense; the first to condense is water, below effective temperatures (\teff) of $\sim$400 K. \ct{Morley14} presented a new grid of model atmospheres for objects from 200--450 K including water ice clouds which become optically thick in Y dwarfs cooler than 350--375 K.

\subsection{Observed Variability in L and T Dwarfs }

Early searches for ultracool dwarf variability focused on the L dwarfs and found evidence for low-amplitude variability \cp[e.g][]{Bailer-Jones01,Gelino02, Clarke08}. A turning point in the field occurred with the discovery of high amplitude variability in the near-infrared in two L/T transition objects \cp{Artigau09, Radigan12}. Today, with a combination of higher precision ground- and space-based data, the study of variability in brown dwarfs is reaching maturity. Brown dwarfs of spectral types from L to Y have been observed to be variable using photometry \cp{Artigau09, Radigan12, Gizis13, Biller13} or spectroscopy \cp{Buenzli12, Apai13, Buenzli14, Burgasser14}. The shape of observed light curves is not always sinusoidal and repeated observations days apart show evolution \cp{Artigau09, Gillon13, Biller13}. 


Different wavelengths probe different layers of a brown dwarf; by observing spectral variability we can understand both the causes of variability and the vertical structure. For example, \ct{Buenzli12} observed phase lags between variability at different wavelengths and found a correlation between pressure probed and phase lag. The complex, evolving nature of variability suggests that many physical processes are involved.

\subsection{Two mechanisms that cause variability}

There are two classes of physical processes that would cause variability in T and Y dwarfs. One class is heterogenous opacity sources in the atmosphere, either caused by non-uniform chemical abundances or cloud cover. We will focus on the role of clouds. The second class is non-uniform temperature structure, either ``hot spots'' or ``cold spots,'' and may be caused by effects of 3D circulation or radiative interaction between deeper patchy clouds and the overlying atmosphere \cp{Showman13, Robinson14}. Here, we present models in each of these categories and make predictions for photometric and spectroscopic variability.  

\vspace{5mm}

\section{Variability from Patchy Clouds}

If one hemisphere has a larger fraction of the surface covered by clouds than the other, as the brown dwarf rotates, the cloudier hemisphere comes in and out of view, and we observe variable brightness. 

We estimate the spectral variability using 1D models that include patchy sulfide/salt and water clouds; briefly, these models follow the approach of \ct{Marley10, Morley14}; we calculate flux separately through both a cloudy column and a cloud-free (clear) column and sum these columns together to calculate the total emergent flux. We can change the cloud-covering fraction by varying $h$, the fractional area assumed to be covered in holes: 

\begin{equation}
F_{\rm total} = h F_{\rm clear} + (1-h) F_{\rm cloudy}
\end{equation}
Using this summed flux $F_{\rm total}$ through each atmospheric layer, we iterate to find a solution in radiative--convective equilibrium. Thus the total flux is the area-weighted sum of the flux from the clear and cloudy columns. Neither column alone carries the flux associated with the combined effective temperature.

The cloud properties for water ice and sulfide/salt clouds are presented in \ct{Morley14} and  \ct{Morley12} respectively. The atmosphere models are presented in detail in \ct{Mckay89, Marley96, MM99, Saumon12}. 

\subsection{Partly Cloudy Spectra}

To calculate the pressure--temperature (\emph{P--T}) structures used here, $h$=0.5 (50\% cloudy). However, both hemispheres do not necessarily have the same cloud-covering fraction. When the clouds/holes are distributed non-uniformly, variability will be observed; the hemisphere with more holes is brighter and has a higher apparent \teff. The amplitude of variability is calculated by summing the flux through the clear and cloudy columns in different proportions which must sum to a net cloud-cover of 50\% to match the \emph{P--T} profile. 

One strength of this method is that using a single, global \emph{P--T} profile isolates the effect of the cloud opacity. Furthermore, the entropy deep within the atmosphere's convective zone must meet the interior entropy; a given pressure should be horizontally uniform in temperature. Our method captures that fact, instead of modeling cloudy and clear regions with the same \teff\ but very different internal entropy. This approach implicitly assumes that the columns are interacting with each other dynamically, an assumption that breaks down for very large, hemispheric patches.

 \begin{figure}[t]
 \center    \includegraphics[width=3.75in]{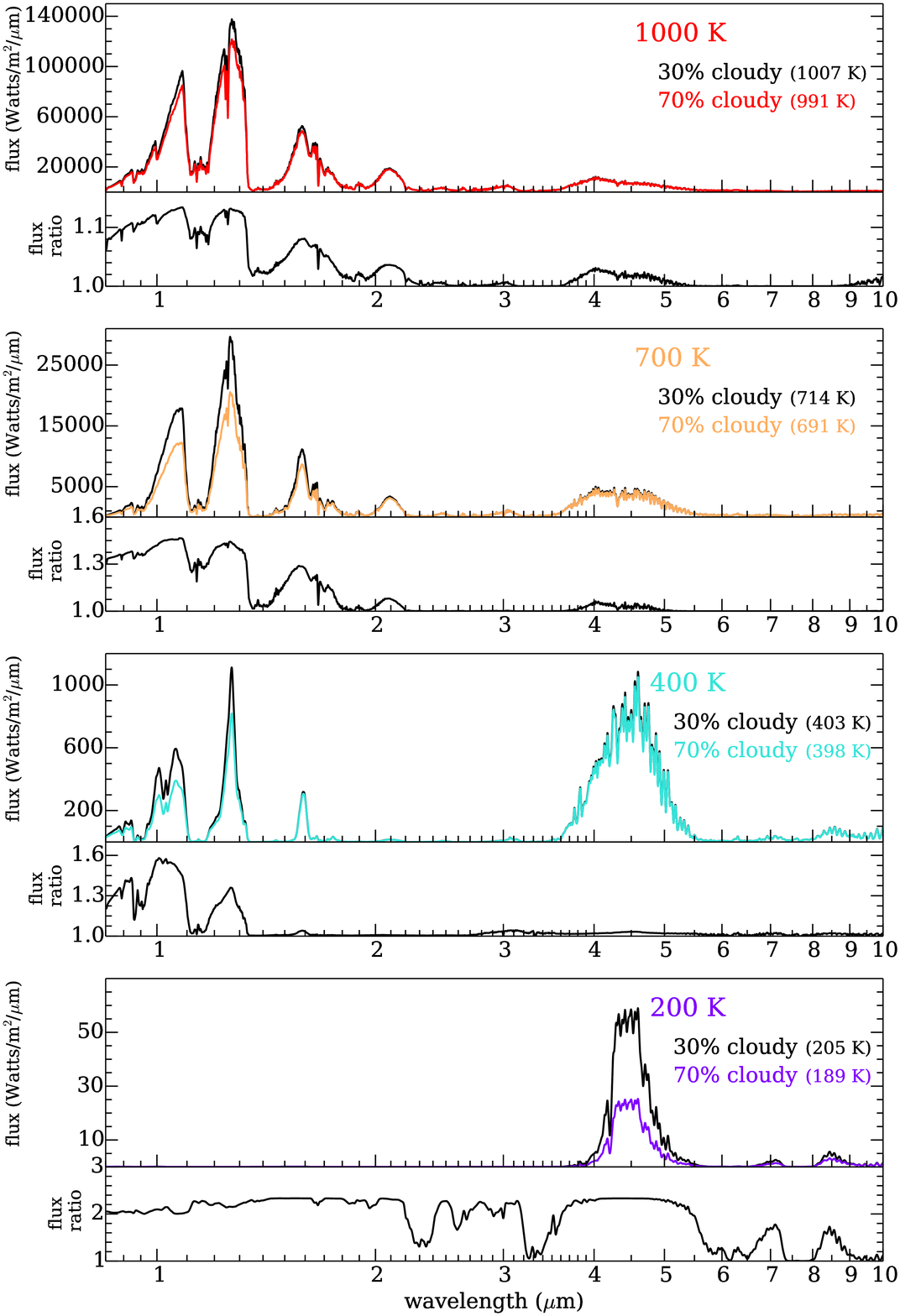}
  \caption{Spectra of partly cloudy models from \teff=1000 K to 200 K. Each pair of panels shows a different summed \teff. Spectra for each \teff\ are calculated using a single 50\% cloudy model with the cloud parameter \fsed=5 in radiative--convective equilibrium. The spectra represent two heterogeneous hemispheres of a 50\% cloudy brown dwarf. Apparent \teff\ of each hemisphere is shown in parentheses. The flux ratio (the ratio of the plotted spectra) is shown in the bottom panel of each pair.  }
\label{patchycloud-spectra}
\end{figure}

 \begin{figure}[t]
 \vspace{-0.7cm}
 \center    \includegraphics[width=3.75in]{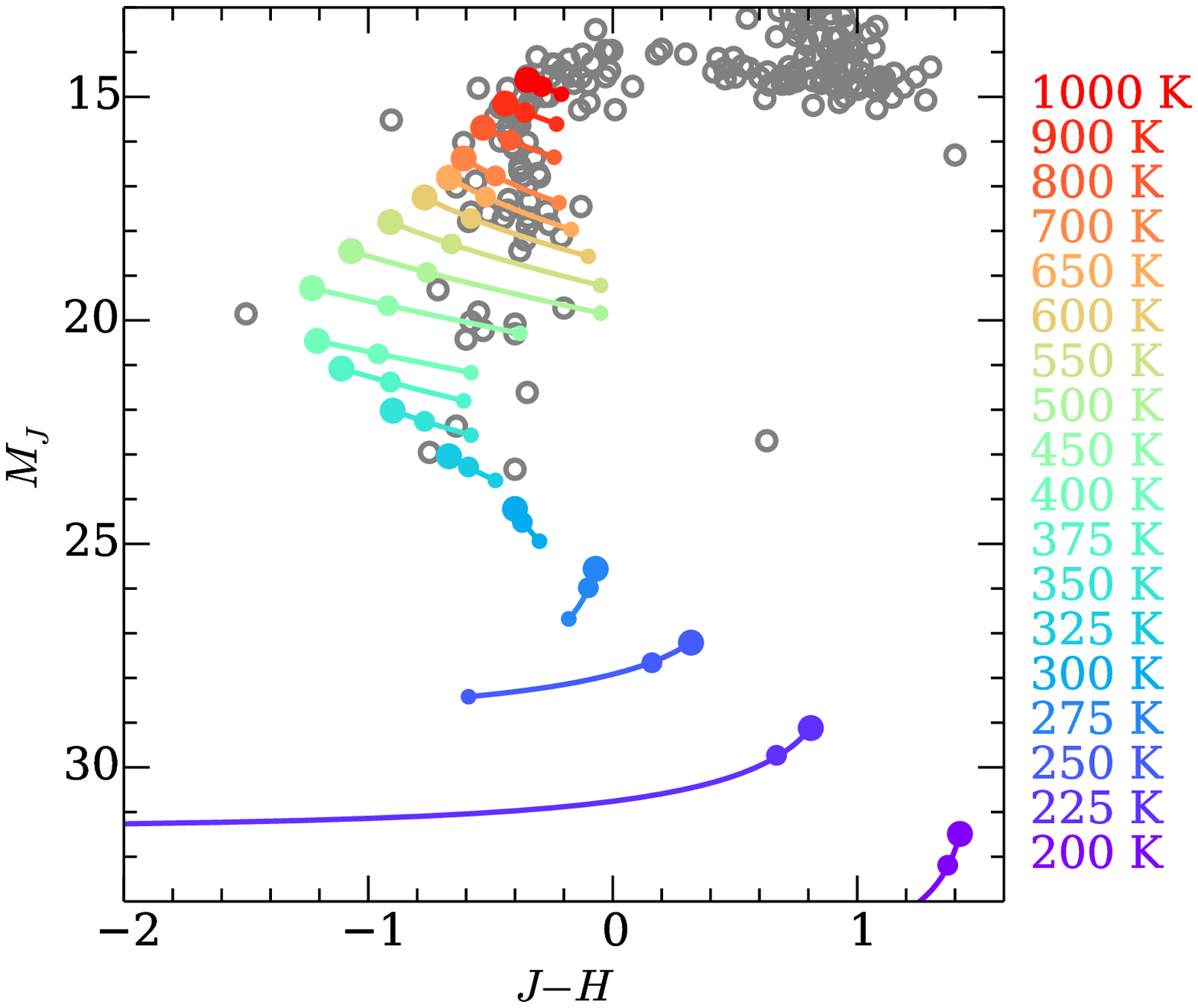}
\label{patchyclouds-cmd-1}
  \vspace{-1.5cm}
\center    \includegraphics[width=3.75in]{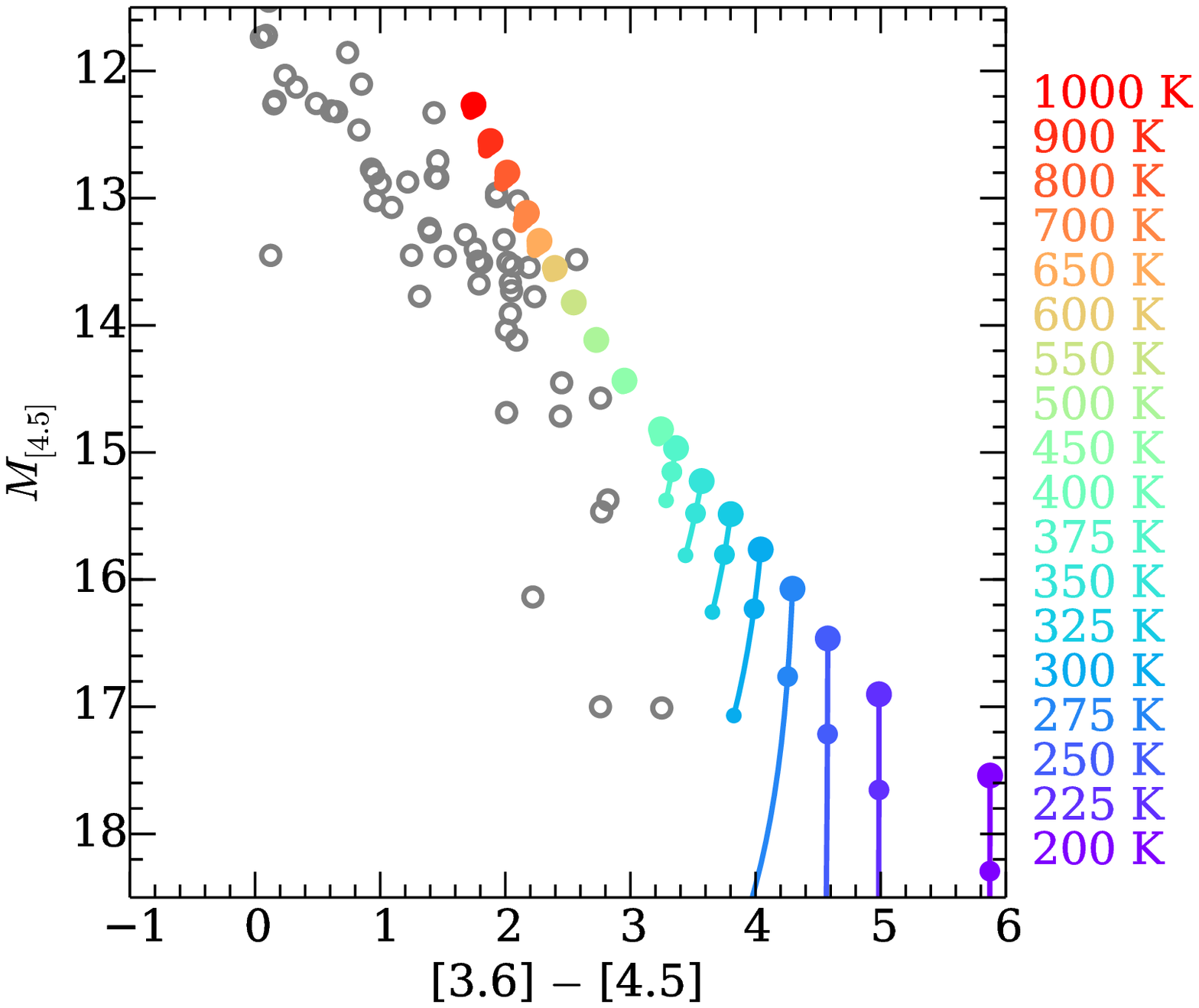}
  \caption{Color--magnitude diagrams for partly cloudy models. The center medium-sized dot represents the 50\% cloudy model in radiative--convective equilibrium. The connected large and small dots show the photometry of the clear and cloudy columns respectively. The \teff\ corresponding to each color is shown on the right of each panel. The observed brown dwarfs with distance measurements are shows as gray open circles \cp{Dupuy12}. The top panel shows $J-H$ vs. $M_J$; the bottom panel shows $[3.6]-$[4.5] vs. $M_{[4.5]}$. }
\label{patchyclouds-cmd}
\end{figure}

Example spectra from \teff=1000 to 200 K are shown in Figure \ref{patchycloud-spectra}. The black lines show the flux emitted from a 30\% cloudy hemisphere; the colored lines show flux emitted from a 70\% cloudy hemisphere. Less flux emerges through the cloudier hemisphere because clouds increase the total opacity. 

The flux ratio between hemispheres is shown in the bottom panels of Figure \ref{patchycloud-spectra}; the flux ratio shows quantitatively the predicted spectral variability. The highest amplitude is within spectral windows, between the major molecular opacity sources in the atmosphere. For \teff=700--1000 K models, the strongest variability is in \emph{Y}, \emph{J} and \emph{H} bands with lower-amplitude variability in \emph{K} band, between 3.6 and 5 \micron, and within the water absorption features. 

In the 400 K model, the variability is largest in \emph{Y} and \emph{J} bands with lower level variability at other wavelengths. Flux at longer wavelengths emerges from higher altitudes than the sulfide and salt clouds, so cloud opacity alone does not change the spectra.

The predicted variability at \teff=200 K looks fundamentally different from the warmer models; this is because by 200 K, the water cloud is thick and dominates the cloud opacity. The flux ratio is nearly uniform from 0.7 to 5.5 \micron, with dips within the major methane absorption features at 2.3 and 3.3 \micron. At this temperature range, significant hemispheric differences in cloud cover cause large amplitude variability at most wavelengths. 

\subsection{Partly Cloudy Color--Magnitude Diagrams}

Model photometry for the partly cloudy models are calculated using radii from the cloud-free \ct{Saumon08} evolution models. The photometry is calculated for the 50\% cloudy converged models and the cloudy and clear columns of each model separately. Two sample color--magnitude diagrams (CMDs) are shown in Figure \ref{patchyclouds-cmd}. The clear, 50\% cloudy, and fully cloudy photometry are shown as large, medium, and small dots connected with a line. 

A near-infrared CMD ($J-H$ vs. $M_J$) is shown in the top panel of Figure \ref{patchyclouds-cmd}. If variability in T and Y dwarfs were due solely to heterogenous clouds, the brown dwarf would move from the center dot along the line that connects the column photometry. For brown dwarfs with \teff$\textgreater$300 K, the object would become redder and somewhat fainter as the cloudier side rotates into view; the sulfide/salt clouds that dominate have the largest impact on \emph{J} (and \emph{Y}) bands. The impact of the sulfide/salt clouds peaks at \teff=500--600 K. 

For brown dwarfs below \teff$\sim$ 300 K, increasing the cloud covering fraction tends to make the brown dwarf bluer in $J-H$. This new behavior is because those objects have thick water ice clouds, which are extremely nongray absorbers. Water ice particles predominantly scatter in \emph{J} band, but absorb more strongly in \emph{H} band and longer wavelengths (see \ct{Morley14}). The water clouds become extremely thick for 200--250 K objects, causing almost all the flux emerging from those objects to emerge through the clear column of the atmosphere; the cloudy point on the CMD becomes extremely blue and faint. 

Likewise, in the mid-infrared CMD ($[3.6]-$[4.5] vs. $M_{[4.5]}$) shown in the bottom panel of Figure \ref{patchyclouds-cmd}, the models separate into two groups. In objects with \teff$\ge$ 400 K, sulfide/salt clouds dominate. However, the sulfide/salt clouds minimally affect the mid-infrared wavelengths (see also Figure \ref{patchycloud-spectra}) so $M_{[4.5]}$ and $M_{[3.6]}$ stay nearly constant. Changes in cloud opacity do not cause significant variability in the mid-late T dwarfs in \emph{Spitzer} observations. In contrast, for models with \teff $\textless$ 400 K, water clouds start to have appreciable optical depth in the mid-infrared where they absorb strongly. The cloudy column becomes fainter in [4.5] and somewhat bluer in [3.6]$-$[4.5]. 

\section{Variability from Hot Spots} \label{hotspots}

Clouds are not the only likely driver of variability; atmospheric dynamics may drive perturbations to the temperature structures. Dynamical effects may create rising and sinking parcels of air on timescales faster than the parcel can equilibrate, causing cold or hot regions. The upper atmosphere may react radiatively to changes in the deep atmosphere, such as heterogenous cloud opacity or dynamically driven perturbations.  \ct{Robinson14} show that temperature perturbations at $\sim$10 bar can be communicated to the overlying parts of the atmosphere through radiative heating, potentially generating complex time-dependent behaviors, including phase shifts.

\begin{figure}[t]
 \center    \includegraphics[width=3.75in]{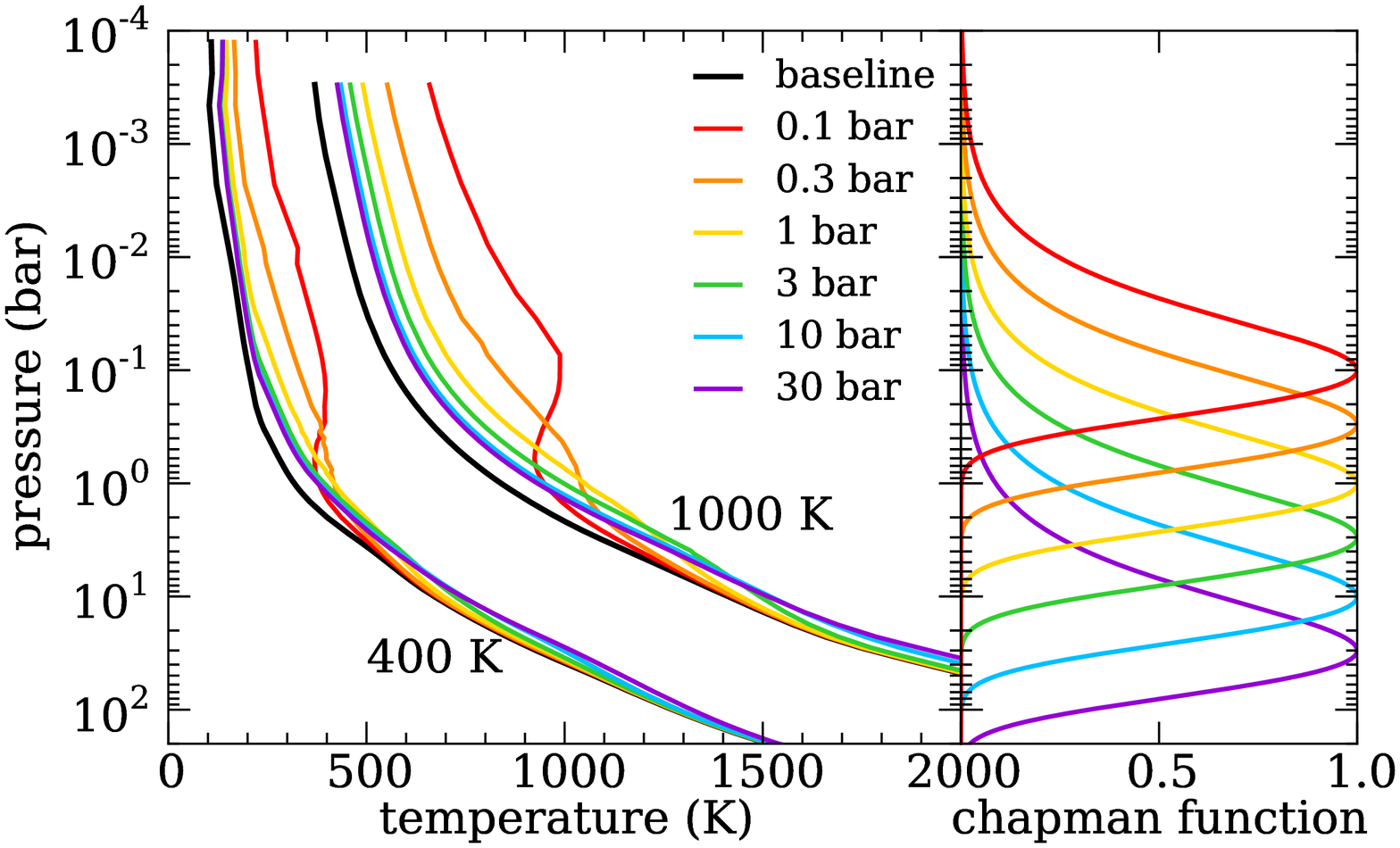}
 \vspace{-1cm}
 \center    \includegraphics[width=3.75in]{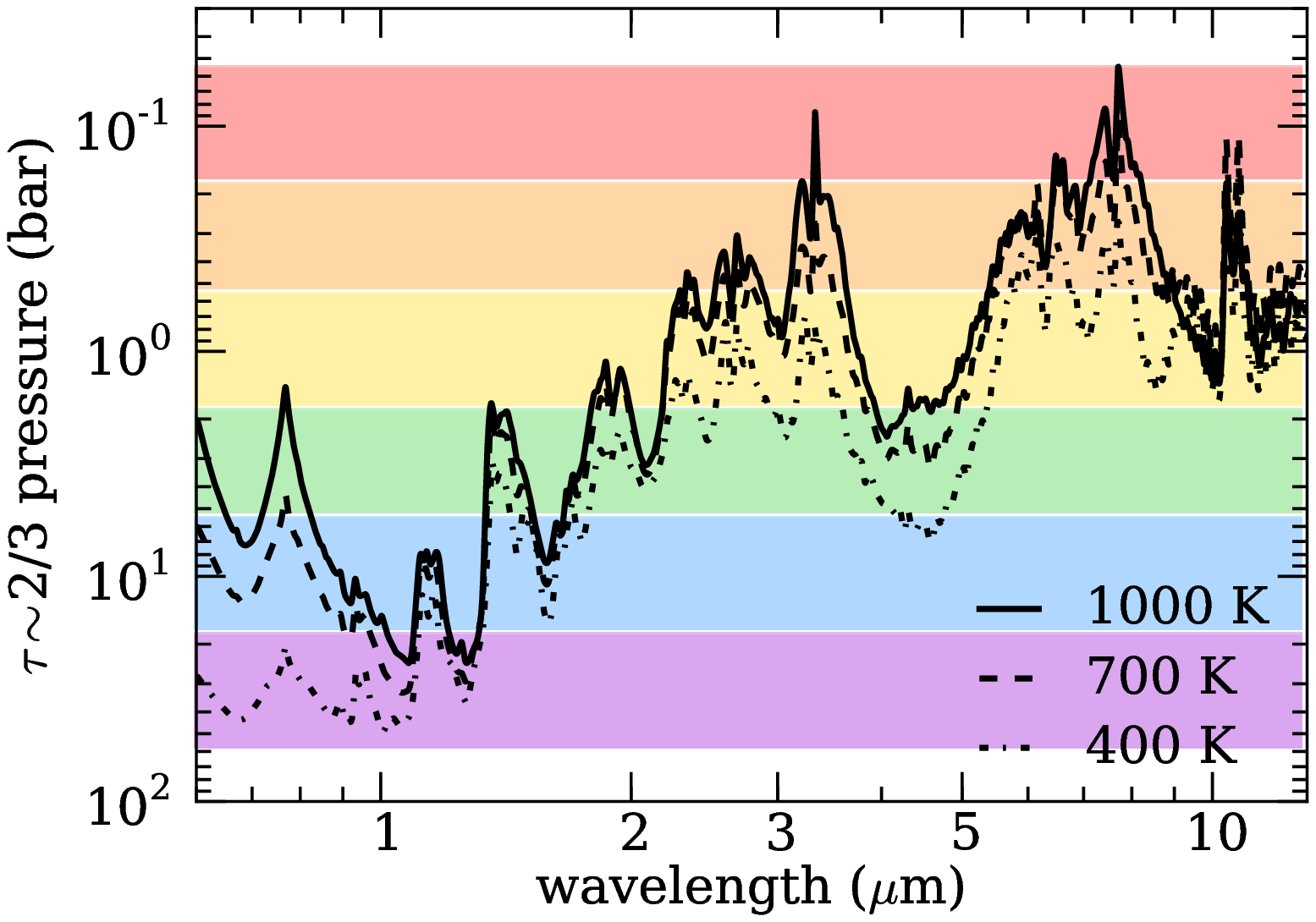}
  \caption{Top panel: Perturbed and unpeturbed pressure--temperature profiles (left) and heating functions (right). The baseline models at \teff=400 and 1000 K are shown in black. The colored lines show models with \emph{P--T} profiles calculated including an additional energy source with the shape of the heating function in the right panel. Bottom panel: the `pressure spectrum' of models with \teff=1000, 700, and 400 K. The colored bars show the same pressure levels as the top panel, at which the perturbations to the profiles are centered. The black lines show the approximate location of the $\tau=2/3$ pressure level as a function of wavelength for the unperturbed models. }
\label{pt-profile}
\end{figure}

 We incorporate heterogeneous temperature profiles by adding energy at specified pressure levels of static cloud-free model atmospheres from 400--1000 K as we calculate the \emph{P--T} structure in radiative--convective equilibrium. The perturbations have the shape of a Chapman function, which is often used to represent heating by incident flux within molecular bands \cp[e.g.][]{ChambHunt, Marley99}. This provides a reasonable approximation of energy added by, e.g., heating from thermal flux from below through holes in the clouds. We use a Chapman function with a width of a single pressure scale height and amplitude to give total emergent flux $F_{\rm new}=1.5F_{\rm baseline}$. We inject energy at pressure levels from 0.1--30 bar. The \emph{P--T} profiles of the warmest and coldest model in the grid (\teff=400 and 1000 K) are shown in the top left panel of Figure \ref{pt-profile}; the location of the heating function is shown in the right panel. The bottom panel of Figure \ref{pt-profile} shows the location of the $\tau=2/3$ pressure level as a function of wavelength; the colored bands indicate the perturbed pressure levels shown in the top panel. 

\subsection{Hot Spot Spectra}

 \begin{figure}[t]
 \center    \includegraphics[width=3.75in]{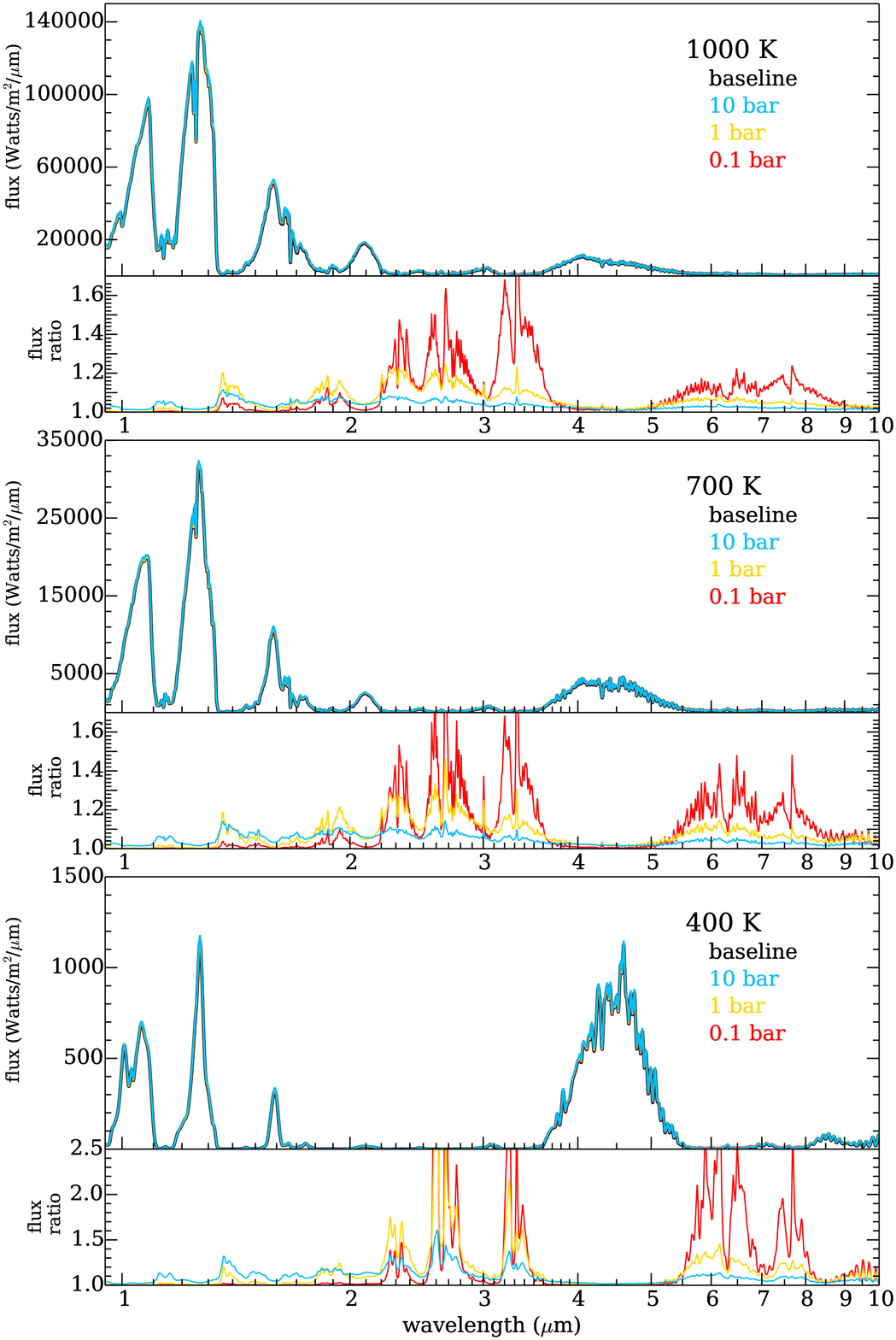}
  \caption{Spectra of models with heated \emph{P--T} profiles from baseline \teff=1000 K to 400 K. Each pair of panels shows a different \teff. The baseline model is shown as a black line. The red, gold, and blue lines show models with 5\% of the surface covered in a hot spot, with heating at 0.1, 1, and 10 bar, respectively. The flux ratio (the ratio of the heated model divided by the baseline model) is shown in the bottom panel of each pair.  }
\label{hot-spot-spectra}
\end{figure}

Representative spectra of models with perturbed \emph{P--T} profiles are shown in Figure \ref{hot-spot-spectra} from \teff=400--1000 K; for each perturbed model, 5\% of the surface is assumed to be covered by the hot spot. 

The flux ratios look quite different from those due to patchy clouds in Figure \ref{patchycloud-spectra}. For these models, the greatest flux ratio is within absorption features instead of within spectral windows. Especially prominent is the methane feature at 3.3 \micron. 

The spectral dependence of variability is controlled by the layer at which the \emph{P--T} profile is perturbed. Heating high in the atmosphere increases flux emerging from higher altitudes, in the mid-infrared. Heating deep within the atmosphere increases flux more uniformly. By observing variability across many wavelengths, we can distinguish between patchy cloud variability and heating at different levels of the atmosphere.

\subsection{Hot Spot Color--Magnitude Diagrams}

\begin{figure}[t]
  \vspace{-0.7cm}
 \center    \includegraphics[width=3.75in]{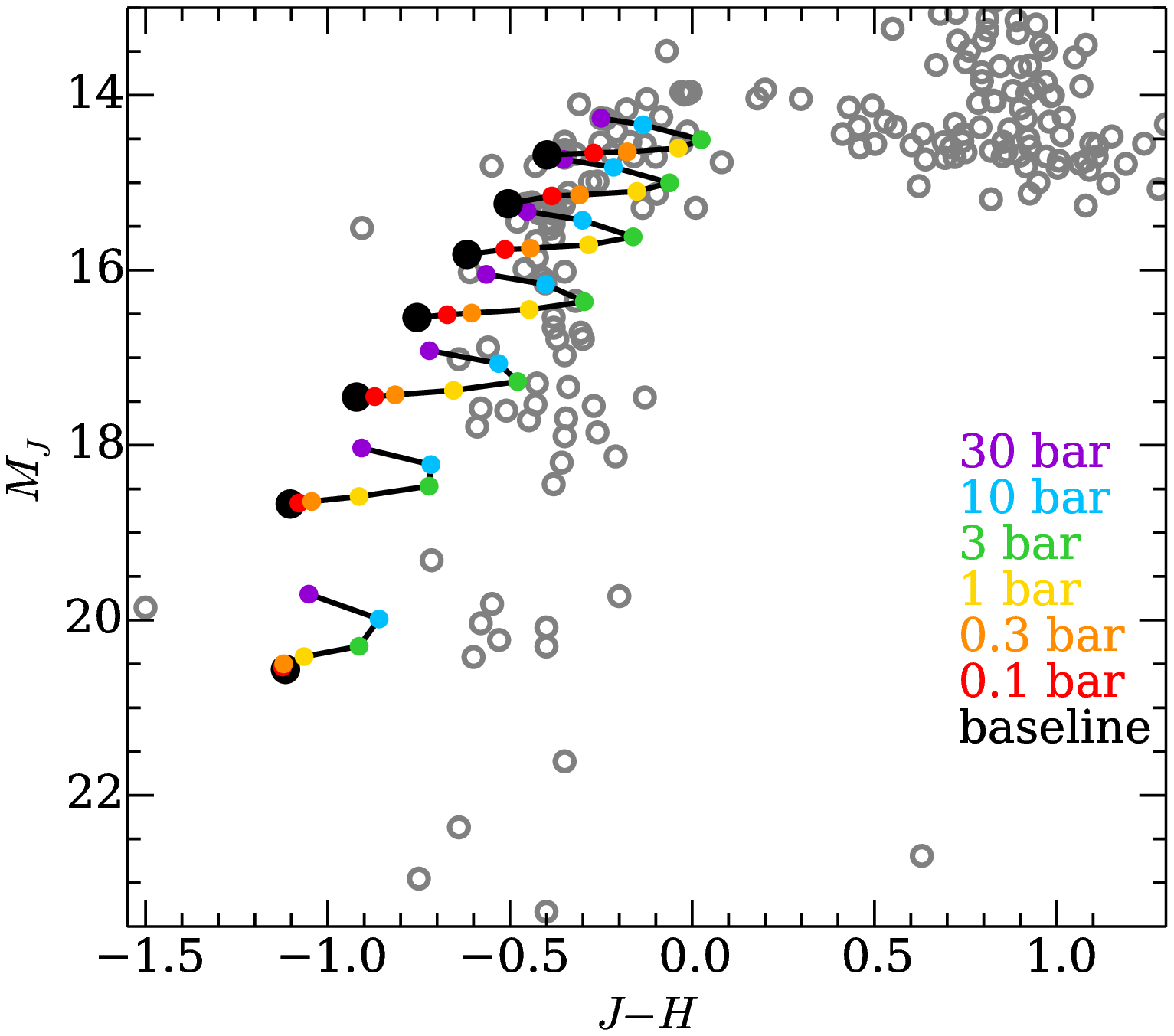}
  \vspace{-1.5cm}
 \center    \includegraphics[width=3.75in]{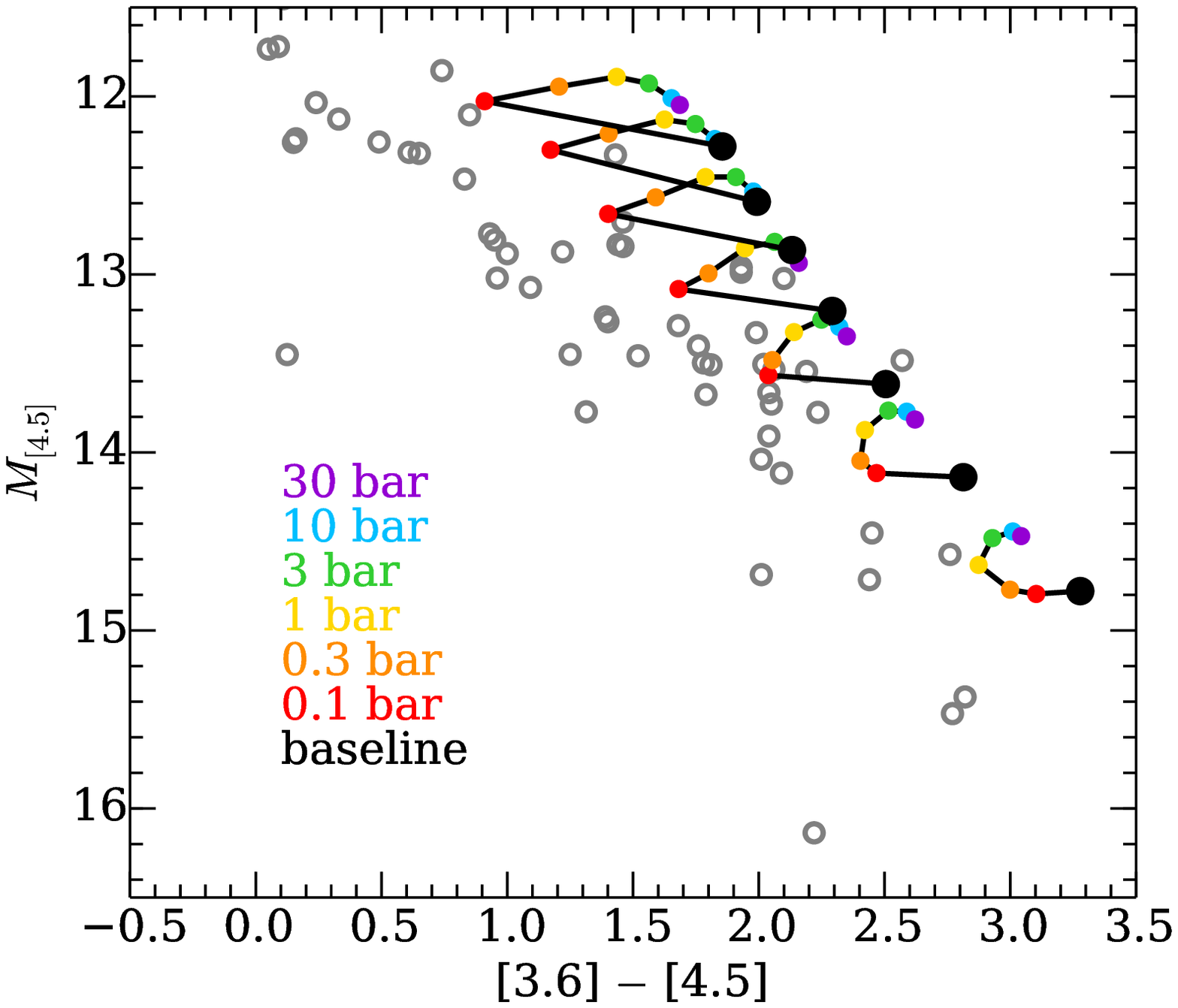}
  \caption{Color--magnitude diagrams for models with perturbed \emph{P--T} profiles. The larger black point shows the photometric point of the `baseline' model for \teff=400--1000 K (in 100 K increments). The colored points show photometry for \emph{P--T} profiles with added energy at each of the specified pressure levels. The observed brown dwarfs with distance measurements are shows as gray open circles \cp{Dupuy12}. The top panel shows $J-H$ vs. $M_J$; the bottom panel shows [3.6]$-$[4.5] vs. $M_{[4.5]}$.}
\label{hot-spot-cmd}
\end{figure}

Near- and mid-infrared CMDs for the models with hot spots are shown in Figure \ref{hot-spot-cmd}. In the top panel ($J-H$ vs. $M_J$), heating high in the atmosphere causes a minimal color and brightness change. The greatest color change occurs when we heat the near-infrared photosphere, around 3--10 bar. Deep heating leads to less chromatic changes.

In the bottom panel ([3.6]$-$[4.5] vs. $M_{[4.5]}$), heating high in the atmosphere causes a very chromatic change, due to significant brightening within the methane band captured in the [3.6] filter. Deeper heating causes less dramatic brightening in both \emph{Spitzer} filters.

\section{Discussion}

\subsection{Simultaneous multi-wavelength observations}

This study suggests that the most illustrative types of observations for understanding the physical processes underlying brown dwarf variability are simultaneous, multi-wavelength observations that probe both inside and outside of molecular absorption features. These measurements are best done from space to avoid the strong molecular absorption of water vapor in Earth's atmosphere. 

Several objects have been observed in such a way to date. Two L/T transition objects, 2M2139 and SIMP0136, were observed using the \emph{Hubble Space Telescope} from 1.1--1.7 \micron, which probes \emph{J} and \emph{H} bands and the water features surrounding those windows. The spectral dependence of the variability observed looks qualitatively similar to the top panel of Figure \ref{patchycloud-spectra}, in which the variability within the spectral windows is larger than the variability within the absorption features. \ct{Buenzli12} present observations of 2MASS J22282889--431026 from partially simultaneous \emph{HST} and \emph{Spitzer} Space Telescope observations. In that object, there are hints that there is larger variability within absorption features: the largest amplitude variability (5.3$\pm$0.6\%) is measured in the 1.35--1.43 \micron\ range. However the other absorption features show similar amplitude variability ($\sim$2\%) as the spectral windows. 

\subsection{Time and length scales for atmospheric heterogeneity}

A number of physical timescales compete in T and Y dwarf atmospheres. The radiative time constant
\begin{equation} 
\tau_{\rm rad}\sim\dfrac{P}{g}\dfrac{c_P}{4\sigma T^3}
\end{equation}
describes the relaxation timescale towards radiative equilibrium following a temperature perturbation \cp{GoodyYung, Fortney08a}. In mid T photospheres, $\tau_{\rm rad}\sim$1--10 hours, increasing to $\tau_{\rm rad}\sim$100 hours for Y dwarf photospheres. The timescale for mixing in convective regions can be approximated using mixing length theory; the mixing timescale is 1--2 orders of magnitude faster than $\tau_{\rm rad}$. The timescale for mixing in radiative regions is  more uncertain and controlled by the interaction of the stable upper atmosphere with the turbulent convective zone, which  generates a wide spectrum of atmospheric waves including gravity waves and Rossby waves \cp{Freytag10, Showman13}. Analytical estimations in \ct{Showman13} suggest that typical timescales for parcels of air to rise or fall one scale height are tens to hundreds of hours. The timescale for radiative relaxation and vertical advection are comparable, creating a complex interplay between atmospheric dynamics and radiative feedback. In addition, the condensation timescale for $\sim$5 \micron\ \nas\ particles \cp[][equation 1]{Carlson88} is of the same order of magnitude. Cool brown dwarfs likely have heterogeneous atmospheres in which rising and falling parcels of air move vapor which condenses on comparable timescales to both the motion and radiative cooling.    

It is challenging to estimate the spatial scales of these heterogeneities from models without better understanding the horizontal wind speeds of brown dwarfs. The sizes of jets in the solar system giant planets generally scale with the Rhines scale, $L_{\rm Rh}\sim(U/2\Omega R \cos\phi)^{1/2}$ where $U$ is wind speed, $R$ is the radius, $\Omega$ is 2$\pi/P$, $P$ is the rotation period, and $\phi$ is the latitude \cp{Rhines70, Showman08b}. \ct{Showman13} estimate a typical brown dwarf Rhines scale to be 10,000--20,000 km, or roughly 5-10\% of a hemisphere, with typical temperature perturbations on isobars of 5--50 K, even ignoring the effect of heterogeneous clouds. Cloud opacity may increase the apparent $T_{\rm bright}$ differences. For example, the 5 \micron\ hot spots on Jupiter are observed to have a $\sim$50 K difference in $T_{\rm bright}$ due to non-uniform cloud and gas opacity \cp{Carlson92}.

\subsection{Role of high resolution spectral mapping}

High resolution Doppler spectral mapping has been used by \ct{Crossfield14} to create a brightness map of the surface of the nearby brown dwarf Luhman 16B. Such techniques are currently limited to the brightest brown dwarfs. Although powerful, these techniques probe limited wavelength ranges and thus a limited pressure level in the atmosphere; the generated map is a map only of that particular level. In addition, they are most sensitive to a single molecule (e.g. CO), which means that abundance variations could also cause the observed brightness map. This technique is most powerful when combined with the simultaneous multi-wavelength observations that probe a much larger part of the brown dwarf atmosphere and are affected by a number of absorbing species. 

\subsection{Giant Planets: Effect of gravity on variability}

Further study is necessary to understand the effect of gravity on spectroscopic variability. There is evidence that warm planet-mass objects of a given temperature have thicker clouds than higher mass brown dwarfs \cp{Currie11, Barman11b, Madhu11c, Liu13}. \ct{Marley12} suggest that the apparent thickness naturally emerges as a result of low gravity and that the process that may break up clouds at the L/T transition may be gravity-dependent, causing lower-gravity objects to become mostly clear T dwarf-like objects at lower \teff. The interplay of gravity, \teff, and atmospheric dynamics is currently not well-understood. Observations of variability in planets or low-gravity brown dwarfs and comparisons with higher mass brown dwarfs could shed light on these physical processes. \ct{Kostov13} conclude that 1\% amplitude photometric variability will be detectable with next-generation AO systemics such as the Gemini Planet Imager, while the James Webb Space Telescope and 30-meter class telescopes will provide spectral mapping data. \ct{Snellen14} suggest that using 30-m class telescopes, high-resolution Doppler mapping will be possible for the brightest directly imaged planets such as beta Pictoris b.

\section{Summary}

We present models of brown dwarfs that include two drivers of spectroscopic variability: patchy clouds and hot spots. We find that the two mechanisms have different spectral dependence, with patchy clouds driving the highest amplitude variability within spectral windows and hot spots driving larger variability within absorption features. 

From patchy sulfide and salt clouds in objects over 300 K, the largest amplitude variability is within near-infrared opacity windows; objects become redder in near-infrared colors (e.g. $J-H$) as the cloudy side rotates into view. Variability in the mid-infrared would be significantly smaller. In objects below 375 K, water clouds are important and affect the spectrum strongly in the mid-infrared, especially within the 4.5 \micron\ window. Water clouds cause a blueward shift in the near-infrared ($J-H$) as the cloudier side rotates into view because water clouds do not absorb as strongly in \emph{J} as they do in \emph{H} or \emph{K}. 

From heating in the atmosphere at different pressure levels, the spectrum changes predominantly within the absorption features. The highest amplitude variability occurs at the wavelengths that probe the pressure levels where the perturbation is centered. For example, the methane feature at 3.3 \micron\ probes high in the atmosphere; heating at high altitudes ($\sim$0.1 bar) causes the highest amplitude variability within that feature. Heating deeper within the atmosphere warms the whole atmosphere more uniformly and causes the brown dwarf to look like a warmer object. 

By analyzing simultaneous multi-wavelength spectral variability, we can disentangle the physical processes causing brown dwarf variability. By observing these processes over long time periods for a larger sample of objects, we can study atmospheric dynamics and the evolution of weather on substellar extrasolar objects. 

\acknowledgements
We acknowledge Didier Saumon for providing models and for helpful comments on this paper. We acknowledge the Database of Ultracool Parallaxes maintained by Trent Dupuy. JJF acknowledges the support of NSF grant AST-1312545 and MSM acknowledges the support of the NASA Astrophysics Theory and Origins Programs.

\bibliographystyle{apj}

\end{document}